\documentclass{elsart}
\journal{Chaos, Solitons and Fractals}
\newcommand{\beq}{\begin{equation}}
\newcommand{\eeq}{\end{equation}}
\newcommand{\bea}{\begin{eqnarray}}
\newcommand{\eea}{\end{eqnarray}}
\newcommand{\uvec}[1]{{\bf \hat{#1}}}
\newcommand{\eq}[1]{{(\ref{#1})}}
\usepackage{amsmath}
\usepackage{amssymb}
\usepackage{graphicx}
\begin{document}

\begin{frontmatter}
\title{Spin-transfer torque induced reversal in magnetic domains}
\author[iit]{S.~Murugesh}
\author[bard]{M.~Lakshmanan}
\ead{lakshman@cnld.bdu.ac.in}
\corauth[auth]{Corresponding author. Tel: +91~431~2407093, Fax:+91~431~2407093}
 
\address[iit]{Department of Physics \& Meteorology, IIT-Kharagpur, Kharagpur 
721~302, India}
\address[bard]{Centre for Nonlinear Dynamics, School of Physics, 
Bharathidasan University, Tiruchirappalli 620024, India}

\begin{abstract}
Using the complex stereographic variable representation for the macrospin, 
from a study of the nonlinear dynamics underlying the generalized 
Landau-Lifshitz(LL) equation with Gilbert damping, we 
show that the spin-transfer torque is effectively equivalent to an 
applied magnetic field. We study the macrospin switching on a Stoner particle 
due to spin-transfer torque on application of a spin polarized current. 
We find that the
switching due to spin-transfer torque is a more effective alternative to
switching by an applied external field in the presence of damping. We 
demonstrate numerically that a spin-polarized current in the form of a
short pulse can be effectively employed to achieve the desired macro-spin 
switching.

\end{abstract}

\end{frontmatter}

\section{Introduction}
In recent times the phenomenon of {\it spin-transfer torque} has gained much 
attention in nanoscale ferromagnets\cite{slonc:1996,berg:1996,stiles:2006}. 
{\it Electromigration} refers to the recoil linear momentum imparted on the 
atoms of
a metal or semiconductor as a large current is conducted across. Analogously, 
if the current is spin-polarized, the transfer of a strong current across 
results in a transfer of spin angular momentum to the atoms. 
This has lead to the possibility of current induced switching of magnetization 
in nanoscale ferromagnets. With the success of GMR, this has immense 
application potential in magnetic recording devices such as 
MRAMs\cite{stiles:2006,wolf:2006,nesbet:1998,tsang:1998}. The phenomenon
has been studied in several nanomagnetic pile geometries.
The typical set up consists of a 
nanowire\cite{stiles:2006,kelly:1999,kelly:2003,hoffer:2002,myers:1999,wegrowe:2001}, or 
a spin-valve pillar, consisting of two 
ferromagnetic layers, one a long ferromagnetic {\it pinned} layer, and another 
small ferromagnetic layer or film, separated by a spacer conductor layer 
(see Figure 1). 
The pinned layer acts as a reservoir for spin polarized current which on 
passing 
through the conductor and on to the thin ferromagnetic layer induces an 
effective torque on the spin magnetization in the thin film ferromagnet. 
A number of experiments have been conducted on this geometry and the 
phenomenon has been convincingly confirmed
\cite{kat:2000,gro:2001,kent:2003,bass:2003}. 
Although the microscopic quantum theory of the 
phenomenon is 
fairly well understood, interestingly the behavior of the average spin 
magnetization vector can be described at the semi-classical level by the 
LL equation with an additional term\cite{baz:1998}. 

\begin{figure}[h]
\centering\includegraphics[width=.9\linewidth]{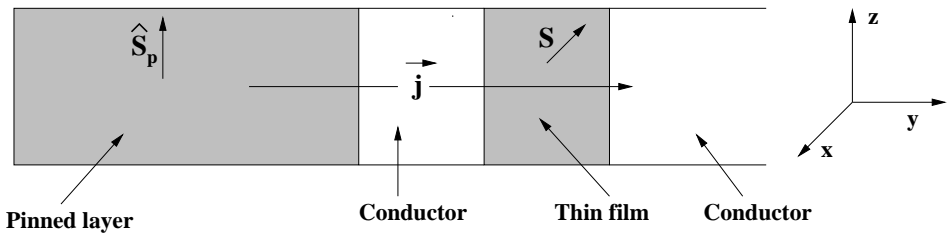}
\caption{A schematic diagram of the spin-valve pillar. A thin film
ferromagnetic layer with magnetization ${\bf S}$ is separated from long 
ferromagnetic layer by a conductor. $\uvec{S}_p$ is the direction of 
magnetization in the pinned region, which also acts as a reservoir for spin 
polarized current.}
\end{figure}

From a different point of view, several studies have focused on  magnetic pulse induced switching of 
the macro-magnetization vector in a thin nanodot under different circumstances
\cite{bauer:2000,gerr:2002,kaka:2002,baz:2004}. Several experimental studies
have also focussed on spin-current induced
switching in the presence of a magnetic field, switching behavior for 
different choices of the angle of the applied field, variation in the 
switching time, etc., 
\cite{kat:2000,bonin:2006,dev:2007,ito:2007,dev:2008,berk:2008}. A numerical 
study on the switching phenomenon induced by a spin current in the presence of 
a magnetic field pulse has also been investigated very recently in 
\cite{pham:2008}. As an extension
to two dimensional spin configurations, the switching behavior on 
a vortex has been  studied in \cite{caputo:2007}. 

In this article, by investigating the nonlinear dynamics underlying the 
generalized Landau-Lifshitz equation with Gilbert damping,  we look at the 
exciting possibility of designing solid state 
memory devices at the nanoscale, wherein memory switching is 
induced using a spin polarized current alone, without the reliance on an 
external magnetic field. We compare earlier studied switching behavior for the 
macro-magnetization vector in a Stoner particle \cite{bauer:2000} in the
presence of an external magnetic field, and the analogous case wherein the
applied field is now replaced by a spin polarized current induced spin-transfer
torque, i.e., with the thin film in the first case replaced by a spin 
valve pillar. It will be shown that a pulse of spin polarized current is
more effective in producing a switching compared to an applied field.  
In doing so we rewrite the system in terms of a complex stereographic variable
instead of the macro-magnetization vector. This brings a significant clarity
in understanding the nonlinear dynamics underlying the macrospin system. 
Namely, it will be shown that, in the complex system, the 
spin-transfer torque is 
effectively an imaginary applied magnetic field. Thus the spin-transfer term
can accomplish the dual task of precession of the magnetization vector and 
dissipation.

The paper is organized as follows: In Section 2 we discuss briefly the 
model system and the associated extended LL equation. In Section 3 we introduce
the stereographic mapping of the constant spin magnetization vector to a 
complex variable, and show that the spin-transfer torque is effectively
an imaginary applied magnetic field. 
In Section 4 we present results from our numerical study on spin-transfer 
torque induced switching phenomenon of the macro-magnetization vector, for a
Stoner particle.
In particular, we study two different geometries for the free layer, namely,
(a) an isotropic sphere and (b) an infinite thin film. In applications to
magnetic recording devices, the typical read/write time period is of the 
order of a few nano seconds. We show that, in order to achieve complete 
switching in these scales, the spin-transfer torque induced by a short pulse 
of sufficient magnitude can be affirmatively employed. We conclude in Section 5
with a discussion of the results and their practical importance. 

\section{The extended LL equation}
The typical set up of the spin-valve pillar consists of a long ferromagnetic 
element, or wire, with
magnetization vector pinned in a direction indicated by $\uvec{S}_p$, as shown
in Figure 1. It
also refers to the direction of spin polarization of the spin current. A free
conduction layer separates the pinned element from the thin ferromagnetic film,
or nanodot, whose average spin magnetization vector 
${\bf S}(t)$ (of constant magnitude $S_0$) is the 
dynamical quantity of interest. The cross sectional dimension of the 
layers range around $70-100 nm$, while the thickness of the conduction layer
is roughly $2-7 nm$\cite{stiles:2006,baz:2004}. The free layer thus acts as the memory unit, separated
from the pinned layer cum reservoir by the thin conduction layer. It is
well established that
the dynamics of the magnetization vector ${\bf S}$ 
in the film in the semiclassical limit is efficiently described by an extended 
LL equation\cite{baz:1998}. If $\uvec{m}(=\{m_1,m_2,m_3\}= {\bf S}/S_0)$ is the unit vector 
in the direction of ${\bf S}$, then
\beq\label{ll}
\frac{d\uvec{m}}{dt} = -\gamma\uvec{m}\times \vec{H}_{eff} 
+ \lambda\uvec{m}\times\frac{d\uvec{m}}{dt}
- \gamma a g(P,\uvec{m}\cdot\uvec{S}_p)\uvec{m}\times(\uvec{m}\times\uvec{S}_p),
\eeq
\beq\label{a}
~~ a \equiv \frac{\hslash Aj}{2S_0Ve}.
\eeq
Here, $\gamma$ is the gyromagnetic ratio $(=0.0176~Oe^{-1}ns^{-1})$ and $S_0$ 
is the saturation magnetization (Henceforth we shall assume $4\pi S_0=8400$,
the saturation magnetization value for permalloy).
The second term in \eq{ll} is the phenomenological dissipation term due to 
Gilbert\cite{gilb:2004} with damping coefficient $\lambda$. The last term is 
the extension to the LL equation effecting the spin-transfer torque, where $A$ 
is the area of cross section, $j$ is the current density, and $V$ is the volume
of the pinned layer. $'a'$, as defined in \eq{a}, has the dimension of $Oe$, 
and is proportional to the current density $j$. 
$g(P,\uvec{m}\cdot\uvec{S}_P)$ is given by
\beq
g(P,\uvec{m}\cdot\uvec{S}_p) = \frac{1}{f(P)(3+\uvec{m}\cdot\uvec{S}_p)-4}~;
~~f(P) = \frac{(1+P)^3}{(4P^{3/2})},
\eeq
where $f(P)$ is the polarization factor introduced by 
Slonczewski \cite{slonc:1996}, and $P (0\le P\le 1)$ is the degree of 
polarization of the pinned ferromagnetic layer. For simplicity, we take this 
factor $g$ to be a constant throughout, and equal to 1.
$\vec{H}_{eff}$ is the effective field acting on the spin vector 
due to exchange interaction, anisotropy, demagnetization and applied 
fields: 
\beq\label{hef}
\vec{H}_{eff} = \vec{H}_{exchange} + \vec{H}_{anisotropy} 
+ \vec{H}_{demagnetization} + \vec{H}_{applied},
\eeq
where
\beq
\vec{H}_{exchange}= D\nabla^2\uvec{m},
\eeq
\beq
\vec{H}_{anisotropy} = \kappa (\uvec{m}\cdot\uvec{e}_{\parallel})\uvec{e}_{\parallel},
\eeq
\beq\label{dem}
\nabla\cdot\vec{H}_{demagnetization} = -4\pi S_0\nabla\cdot\uvec{m}.
\eeq
Here, $\kappa$ is the strength of the anisotropy field. $\uvec{e}_{\parallel}$ 
refers to the direction of (uniaxial) anisotropy, In what
follows we shall only consider homogeneous spin states on the ferromagnetic
film. This leaves the exchange interaction term in \eq{hef} redundant, or
$D=0$, while \eq{dem} for $\vec{H}_{demagnetization}$ is readily solved to give
\beq
\vec{H}_{demagnetization} = -4\pi S_0(N_1m_1\uvec{x} + N_2m_2\uvec{y} + 
N_3m_3\uvec{z}),
\eeq
where $N_i, i=1,2,3$ are constants with $N_1+N_2+N_3=1$, and 
$\{\uvec{x},\uvec{y},\uvec{z}\}$ are the orthonormal unit vectors. 
Equation \eq{ll} now reduces to a dynamical equation for a representative 
macro-magnetization vector $\uvec{m}$. 

In this article we shall be concerned with switching behavior in the film 
purely induced by the spin-transfer torque term, and compare the results with 
earlier studies on switching due to an applied field \cite{bauer:2000} in the 
presence of dissipation. Consequently, it will 
be assumed that $\vec{H}_{applied}=0$ in our analysis. 

\section{Complex representation using stereographic variable}
It proves illuminating to rewrite \eq{ll} using the complex 
stereographic variable $\Omega$ defined as\cite{naka:1984,kosaka:2005}
\beq\label{stereo}
\Omega \equiv \frac{m_1+im_2}{1+m_3},
\eeq
so that
\beq\label{m}
m_1 = \frac{\Omega + \bar{\Omega}}{1+|\Omega|^2}~;~~
m_2 = -i\frac{(\Omega - \bar{\Omega})}{1+|\Omega|^2}~;~~
m_3 = \frac{1-|\Omega|^2}{1+|\Omega|^2}.
\eeq
For the spin valve system, the direction 
of polarization of the spin-polarized current $\uvec{S}_p$ remains
a constant. Without loss of generality, we chose this to be the direction
$\uvec{z}$ in the internal spin space, i.e., $\uvec{S}_p = \uvec{z}$. 
As mentioned in Sec. 2, we disregard the exchange term. However, for the 
purpose of illustration, we choose $\vec{H}_{applied}= \{0,0,h_{a3}\}$ for the
moment but take $h_{a3}=0$ in the later sections. Defining
\beq\label{ep}
\uvec{e}_{\parallel} = \{\sin\theta_\parallel\cos\phi_\parallel,
\sin\theta_\parallel\sin\phi_\parallel,\cos\theta_\parallel\}
\eeq
and upon using \eq{stereo} in \eq{ll}, we get 
\bea\label{llo}
 (1-i\lambda)\dot{\Omega} = 
-\gamma(a-i h_{a3})\Omega 
+ im_{\parallel}\kappa\gamma\big{[}\cos\theta_\parallel\Omega 
-\frac{1}{2}\sin\theta_\parallel(\e^{i\phi_\parallel} - \nonumber\\
\Omega^2\e^{-i\phi_\parallel})\big{]}
-\frac{i\gamma4\pi~S_0}{(1+|\Omega|^2)}\big{[}N_3(1-|\Omega|^2)\Omega 
-\frac{N_1}{2}(1-\Omega^2-|\Omega|^2)\Omega\nonumber\\
-\frac{N_2}{2}(1+\Omega^2-|\Omega|^2)\Omega 
- \frac{(N_1-N_2)}{2}\bar{\Omega}\big{]},
\eea
where $m_{\parallel} = \uvec{m}\cdot\uvec{e}_{\parallel}$. Using \eq{m} and 
\eq{ep}, $m_{\parallel}$, and thus \eq{llo}, can be written entirely in terms 
of $\Omega$. 

It is interesting to note that in this representation the spin-transfer torque 
(proportional to the parameter $a$) appears only in
the first term in the right hand side of \eq{llo} as an addition to the
applied magnetic field $h_{a3}$ but with a prefactor $-i$. Thus the spin
polarization term can be considered as an effective applied magnetic field. 
Letting $\kappa=0$, and $N_1=N_2=N_3$ in \eq{llo}, we have
\beq
(1-i\lambda)\dot{\Omega} = - \gamma(a-ih_{a3})\Omega,
\eeq
which on integration leads to the solution 
\bea
\Omega(t) = \Omega(0)~\exp{(-(a-ih_{a3})\gamma t/(1-i\lambda))}\nonumber
\eea
\beq\label{expo}
= \Omega(0)~\exp{(-\frac{a+\lambda h_{a3}}{1+|\lambda|^2}\gamma t)}
~\exp{(-i\frac{a\lambda-h_{a3}}{1+|\lambda|^2}\gamma t)}.
\eeq 
The first exponent in \eq{expo} describes relaxation, or switching,
while the second term describes precession. 
From the first exponent in \eq{expo}, we note that the time scale
of switching is given by $1/(a+\lambda h_{a3})$. $\lambda$ being small,
this implies that the spin-torque term is more effective in switching the
magnetization vector. 
Further, letting $h_{a3}=0$, we note that in the presence of the damping term 
the spin transfer  produces the dual effect of precession and dissipation. 

To start with we shall analyze the fixed points of the system for the
two cases which we shall be concerned with in this article: (i) the isotropic
sphere characterized by $N_1=N_2=N_3=1/3$, and (ii) an infinite thin film 
characterized by $N_1=0=N_3, N_2=1$. 

(i) First we consider the case when the anisotropy 
field is absent, or $\kappa=0$. From \eq{llo} we have
\bea\label{llf}
(1-i{\lambda})\dot{\Omega} =
-a\gamma\Omega -\frac{i\gamma 4\pi S_0}{1+|\Omega|^2}\big{[}N_3(1-|\Omega|^2)\Omega 
-\frac{N_1}{2}(1-\Omega^2-|\Omega|^2)\Omega\nonumber\\
-\frac{N_2}{2}(1+\Omega^2-|\Omega|^2)\Omega 
- \frac{(N_1-N_2)}{2}\bar{\Omega}\big{]}.
\eea
In the absence of anisotropy ($\kappa=0$), we see
from \eq{llf} that the only fixed point is $\Omega_0 = 0.$
To investigate the stability of this fixed point we expand \eq{llf} up to a 
linear order in perturbation $\delta\Omega$ around $\Omega_0$. This gives
\beq\label{stab}
 (1-i{\lambda})\delta\dot{\Omega} = -a\gamma\delta\Omega 
-i\gamma 4\pi S_0[N_3 - \frac{1}{2}(N_1+N_2)]\delta\Omega 
+i{\gamma 2\pi S_0}(N_1-N_2)\delta\bar{\Omega}.
\eeq
For the isotropic sphere, $N_1=N_2=N_3=1/3$, \eq{stab} reduces to
\beq
(1-i{\lambda})\delta\dot{\Omega} = -a\gamma\delta\Omega.
\eeq
We find the fixed point is stable since $a>0$. For the thin film,
$N_1=0=N_3, N_2 =1$. \eq{stab} reduces to 
\beq\label{stab3}
(1-i{\lambda})\delta\dot{\Omega} = -a\gamma\delta\Omega 
+i\gamma 2\pi S_0\delta\Omega - i\gamma\pi S_0\delta\bar{\Omega}.
\eeq
This may be written as a matrix equation for 
$\Psi\equiv(\delta\Omega~,\delta\bar{\Omega})^T$,
\beq
\dot{\Psi} = {\bf M}\Psi,
\eeq
where ${\bf M}$ is a matrix obtained from \eq{stab3} and its complex 
conjugate, whose determinant and trace are
\beq
|{\bf M}| = \frac{(a^2+3\pi^2S_0^2)\gamma^2}{1+\lambda^2}~;
~~Tr({\bf M}) = \frac{(-2a-4\pi S_0\lambda)\gamma}{1+\lambda^2}.
\eeq
Since $|{\bf M}|$ is positive, the fixed point $\Omega_0=0$ is stable if
$Tr|{\bf M}|<0$, or, $(a+2\pi S_0\lambda)>0$.

The equilibrium point (a), $\Omega_0=0$, corresponds to $\uvec{m} =\uvec{z}$. 
Indeed this holds true
even in the presence of an applied field, though we have little to discuss on 
that scenario here. 

(ii) Next we consider the system with a nonzero anisotropy field in the 
$\uvec{z}$ direction.
\eq{llo} reduces to 
\bea\label{stab2}
 (1-i{\lambda})\dot{\Omega} = -a\gamma\Omega 
+ i\kappa\gamma\frac{(1-|\Omega|^2)}{(1+|\Omega|^2)}\Omega
-\frac{i\gamma 4\pi S_0}{(1+|\Omega|^2)}\big{[}N_3(1-|\Omega|^2)\Omega\nonumber\\
-\frac{N_1}{2}(1-\Omega^2-|\Omega|^2)\Omega
-\frac{N_2}{2}(1+\Omega^2-|\Omega|^2)\Omega
- \frac{(N_1-N_2)}{2}\bar{\Omega}\big{]}.
\eea
Here again the only fixed point is $\Omega_0 = 0.$
As in (i), the stability of the fixed point is studied by expanding \eq{stab2}
about $\Omega_0$ to linear order. Following the same methodology in (i)
we find the criteria for stability of the fixed point for the isotropic sphere 
is $(a+{\lambda}\kappa)>0$, while for the thin film
it is $(a+\lambda(\kappa+2\pi S_0))>0$.

(iii) With nonzero $\kappa$ in an arbitrary direction the fixed point in 
general moves away from $\uvec{z}$. 

Finally, it is also of interest to note that a sufficiently large current 
leads to spin wave instabilities induced through spin-transfer torque
\cite{poli:2004,poli:2006}. In the present investigation, however, we have 
assumed homogeneous magnetization over the free layer, thus ruling out such
spin wave instabilities. Recently we have investigated spin wave 
instabilities of the Suhl type induced by an applied alternating field in thin
film geometries using stereographic representation\cite{kosaka:2005}. It will 
be interesting to investigate the role of a spin-torque on such instabilities 
in the spin valve geometry using this formulation. This will be pursued 
separately.

\section{Spin-transfer torque induced switching}
We now look at the interesting possibility of effecting complete switching of 
the magnetization
using spin-transfer torque induced by a spin current. Numerical studies on 
switching effected on a Stoner particle by an applied magnetic field,
or in the presence of both a spin-current and applied field,
in the 
presence of dissipation and axial anisotropy have been carried out recently
and switching has been demonstrated \cite{bauer:2000,pham:2008}. 
However, the intention here is to induce the same using
currents rather than the applied external fields. Also, achieving such
localized magnetic fields has its technological challenges. 
Spin-transfer torque proves 
to be an ideal alternative to accomplish this task since, as we have pointed 
out above, it can be considered as an effective (albeit complex) magnetic
field. In analogy with ref. \cite{bauer:2000}, where switching behavior
due to an applied magnetic field has been studied
we investigate here switching behavior purely due to spin-transfer torque, on
a Stoner particle. 
Numerical results in what follows have been obtained by directly simulating
\eq{llo} and making use of the relations in \eq{m}, for appropriate choice of 
parameters. It should be remembered that 
\eq{llo} is equivalent to \eq{ll}, and so the numerical results have been 
further confirmed by directly numerically integrating \eq{ll} also for the
corresponding parameter values. We consider below two samples differing in 
their shape 
anisotropies, reflected in the values of $(N_1,N_2,N_3)$ in the demagnetization
field: a) isotropic sphere, $N_1=N_2=N_3=1/3$ and b) a thin film 
$N_1=0=N_3, N_2=1$. The spin polarization $\uvec{S}_p$ of the current
is taken to be in the $\uvec{z}$ direction. The initial orientation of
$\uvec{m}$ is taken to be close to $-\uvec{z}$.
In what follows this is taken as  $170^\circ$ from $\uvec{z}$
in the $(z-x)$ plane. The orientation of uniaxial anisotropy 
$\uvec{e}_{\parallel}$ is also taken to be the initial direction of $\uvec{m}$.
With these
specified directions for $\uvec{S}_p$ and $\uvec{e}_\parallel$ the stable
fixed point
is slightly away from $\uvec{z}$, the direction where the magnetization 
$\uvec{m}$ is expected to switch in time. A small damping is assumed, with 
$\lambda = 0.008$. The magnitude
of anisotropy $\kappa$ is taken  to be $45~Oe$. 
As stated earlier, for simplicity we have considered the magnetization to
be homogeneous. 

\subsection{Isotropic sphere}\label{is}
It is instructive to start by investigating the isotropic sphere, which is
characterized by the demagnetization
field with $N_1=N_2=N_3=1/3$. With these values for $(N_1,N_2,N_3)$, \eq{llo} 
reduces to 
\bea\label{lli}
(1-i{\lambda})\dot{\Omega} = -a\gamma\Omega 
+ im_{\parallel}\kappa\gamma\big{[}\cos\theta_\parallel\Omega 
-\frac{1}{2}\sin\theta_\parallel(\e^{i\phi_\parallel}-\Omega^2\e^{-i\phi_\parallel})\big{]}
\eea
A constant current of $a=10~Oe$ is assumed. Using \eq{a}, for typical 
dimensions, this equals a current density of the order $10^8 A/cm^2$. 
We notice that for the isotropic sample the demagnetization field does not play
any role in the dynamics of the magnetization vector. 
In the absence of anisotropy and damping the spin-transfer torque term
leads to a rapid switching of ${\bf S}$ to the $\uvec{z}$ 
direction. This is evident from \eq{lli}, which becomes
\beq
\dot{\Omega} = -a\gamma\Omega,
\eeq
with the solution $\Omega=\Omega_0~\e^{-a\gamma t}$, and the time scale for 
switching is given by $1/a\gamma$.
Figure 2.a shows the trajectory traced out by the magnetization vector 
${\bf S}$, for 
5 ns, initially close to the $-\uvec{z}$ direction, switching to the 
$\uvec{z}$ direction. Figure 2.b depicts the dynamics with anisotropy but no
damping, all other parameters remaining same. While the same 
switching is achieved, this is more smoother due to the accompanying 
precessional motion. 
Note that with nonzero anisotropy, $\uvec{z}$ is not the 
fixed point any more. The dynamics with damping but no anisotropy (Figure 2.c)
resembles Figure 2.a, while Figure 2.d shows the dynamics
with both anisotropy and damping. 
\begin{figure}[h]
\centering\includegraphics[width=1.07\linewidth]{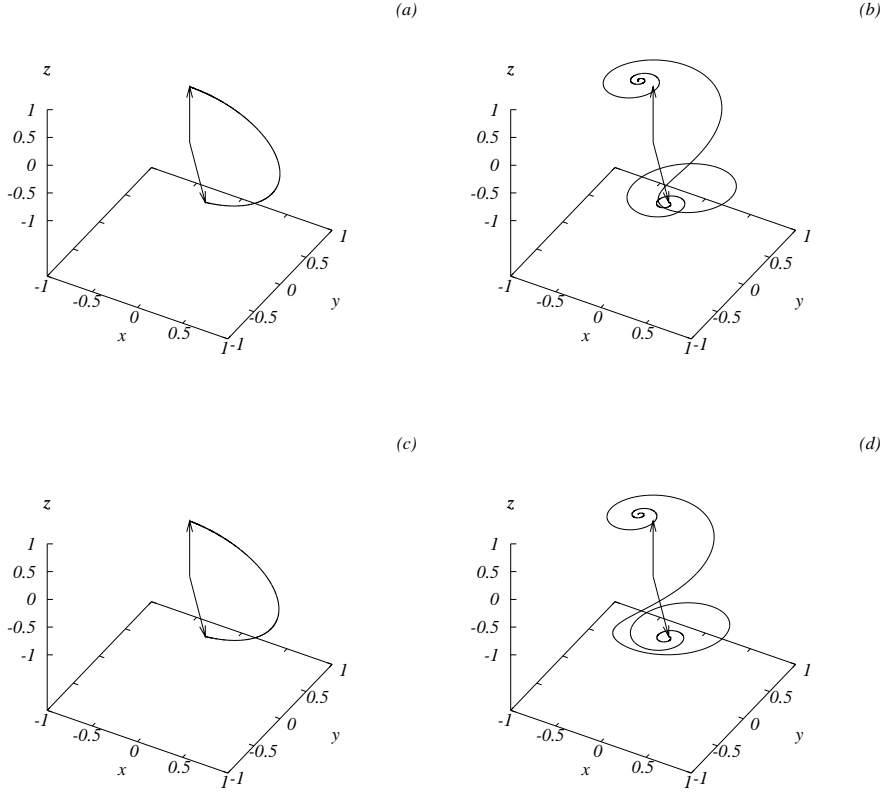}
\caption{Trajectory of the magnetization vector ${\bf m}$, obtained by
simulating \eq{llo} for the isotropic sphere ($N_1=N_2=N_3=1/3$), and using 
the relations in \eq{m}, for 
$a=10~Oe$ (a) without anisotropy and damping, (b) with 
anisotropy but no damping, (c) without anisotropy but nonzero
damping and (d) with both anisotropy and damping nonzero. The results have 
also been confirmed by numerically integrating \eq{ll}. The arrows point
in the initial orientation (close to $-\uvec{z}$) and the direction of the 
spin current $\uvec{z}$. Evolution shown is for a period of $5 ns$. Note 
that the final orientation is not exactly $\uvec{z}$ in the case of nonzero 
anisotropy ((b) and (d)).}
\end{figure}

It may be noticed that Figures 2.c and 2.d resemble qualitatively Figures 2.a 
and 2.b, respectively, while differing mainly in the
time taken for the switching. It is also noticed that switching in the
absence of anisotropy is faster. Precession assisted switching has been the
favored recording process in magnetic memory devices, as it helps in keeping
the exchange interaction at a minimum\cite{gerr:2002,kaka:2002}. The sudden 
switching noticed in the absence of anisotropy essentially refers to 
a momentary collapse of order in the magnetic media. This can possibly lead to 
strong exchange energy and a breakdown of our assumption regarding homogeneity 
of the magnetization field. However, such rapid quenching assisted by short 
high intensity  magnetic pulses has in fact been achieved experimentally 
\cite{tudo:2004}.

A comparison with reference \cite{bauer:2000} is in order. There it was 
noted that with an applied magnetic field, instead of a spin torque, a 
precession assisted switching was possible only in the presence
of a damping term. In Section 3 we pointed out how the spin transfer torque
achieves both precession and damping. Consequently, all four scenarios depicted
in Figure 2 show switching of the magnetization vector without any applied
magnetic field. 
\subsection{Infinite thin film}\label{tf}
Next we consider an infinite thin film, whose demagnetization field is given
by $N_1=0=N_3$ and $N_2=1$. With these values \eq{llo} becomes 
\bea
 (1-i{\lambda})\dot{\Omega} = -a\gamma\Omega 
+ im_{\parallel}\kappa\gamma\big{[}\cos\theta_\parallel\Omega 
-\frac{1}{2}\sin\theta_\parallel(\e^{i\phi_\parallel}
-\Omega^2\e^{-i\phi_\parallel})\big{]}\nonumber\\
- i \gamma 4\pi S_0\Big{(}\frac{1-|\Omega|^2}{1+|\Omega|^2}\Big{)}\Omega. 
\eea
Here again in the absence of anisotropy $\Omega=0$ is the only fixed
point. Thus the spin vector switches to $\uvec{z}$ in the absence of damping 
and anisotropy (Figure 3a). In order to achieve this in a time scale of
$5 ns$, we find that the value of $a$ has to be of order $50~Oe$. 
Again the behavior is in stark contrast to the case induced purely
by an applied field\cite{bauer:2000}, wherein the spin vector traces out
a distorted precessional trajectory. As in Sec. \ref{is}, the trajectory traced
out in the presence of damping is similar to that without damping (Figure 3c). 
The corresponding
trajectories traced out in  the presence of anisotropy are shown in 
Figures 3b and 3d.

\begin{figure}[h]
\centering\includegraphics[width=1.07\linewidth]{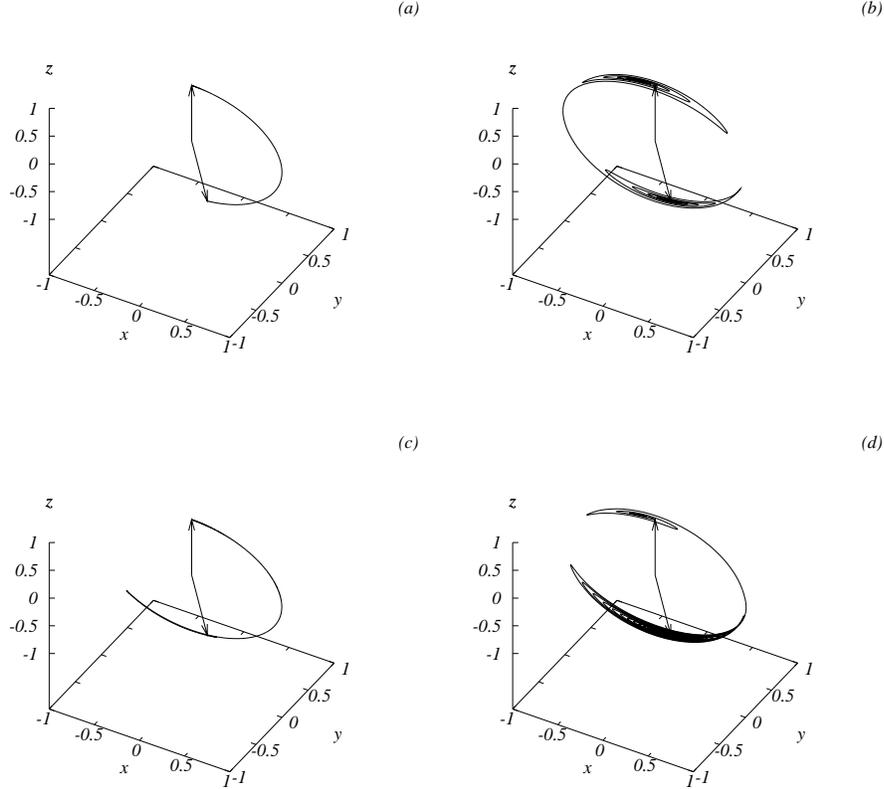}
\caption{Trajectory of the magnetization vector ${\bf S}$ in a 
period of $5 ns$, obtained as earlier by numerically simulating \eq{llo},
and also confirming with \eq{ll},
with demagnetization field with by $N_1=0=N_3$, 
$N_2=1$ and $a=50~Oe$, and all the parameter values are as earlier.
As in Figure 1, the $\uvec{S}$ and initial orientation are
indicated by arrows. (a) Without anisotropy or damping, (b) nonzero anisotropy
but zero damping, (c) without anisotropy but nonzero damping and (d) both
anisotropy and damping nonzero. As earlier, in the presence of nonvanishing
anisotropy, the fixed point is not the $\uvec{z}$ axis. }
\end{figure}

\subsection{Switching of magnetization under a pulsed spin-polarized current}
We noticed that in the absence of uniaxial anisotropy, the constant spin
polarized current can effect the desired switching to the orientation of 
$\uvec{S}_p$ (Figure 2). This is indeed the fixed point for the system (with 
no anisotropy). Figure 2 traces the dynamics
of the magnetization vector in a period of 5 ns, in the presence
of a constant spin-polarized current. 
However, for applications in magnetic media 
we choose a spin-polarized current {\it pulse} of the form shown in Figure 4. 
It may be recalled here that, as was observed in \ref{tf}, with a spin 
polarized current of sufficient 
magnitude, the switching time can indeed be reduced. We choose a pulse,
polarized as earlier along the $\uvec{z}$ direction, with rise
time and fall time of $1.5 ns$, and a pulse width, defined as the time interval
between half maximum, of $4 ns$. We assume the rise and fall phase of the 
pulse to be of a sinusoidal form, though, except for the smoothness, the 
switching phenomenon is independent of the exact form of the rise or fall 
phase.
\begin{center}
\begin{figure}[h]
\centering\includegraphics[width=.8\linewidth]{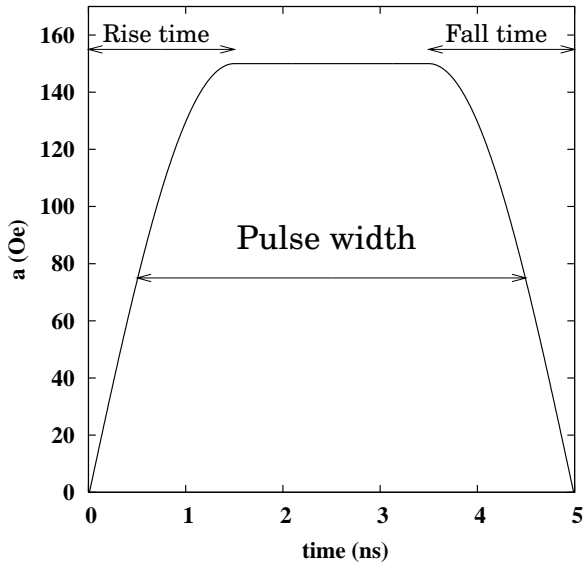}
\caption{Pulse form showing the magnitude of $a$, or effectively the 
spin-polarized current. The rise
and fall phase are assumed to be of a sinusoidal form. The rise and
fall time are taken as $1.5 ns$, and pulse width $4~ns$. The maximum 
magnitude of $a$ is $150~Oe$. }
\end{figure}
\end{center}

In Figures 5 and 6, we show trajectories of the spin vector for a period of
$25~ns$, for the two 
different geometries, the isotropic sphere and a thin film. The action of
the spin torque pulse, as in Figure 4, is confined to the first $5~ns$. 
We notice that, with 
the chosen value of $a$, this time period is enough to effect the switching. 
In the absence of anisotropy, the direction of $\uvec{S}_p$ is the fixed point.
Thus a pulse of sufficient magnitude can effect a switching in the desired 
time scale of $5 ns$. From our numerical study we find that in order for this
to happen, the value of $a$ has to be of order $150~Oe$, or, from \eq{a}, a 
current  density of order $10^9A/cm^2$, a magnitude achievable experimentally
(see for example \cite{tsoi:1998}). Comparing with sections \ref{is} and 
\ref{tf}, we note that the extra one order of magnitude in current density 
is required due to the
duration of the rise and fall phases of the pulse in Figure(4). Here again 
we contrast the trajectories with those induced by an applied magnetic field
\cite{bauer:2000}, where the switching could be achieved only in the
presence of a uniaxial anisotropy.

In Figure 5b for the isotropic sphere with nonzero crystal field anisotropy, 
we notice that 
the spin vector switches to the fixed point near $\uvec{z}$ axis in the first
$5~ns$. However the magnetization vector precesses around $\uvec{z}$ after the 
pulse has been turned off. This is because in the absence of the spin-torque
term, the fixed point is along $\uvec{e}_\parallel$, the direction of uniaxial 
anisotropy. Due to the nonzero damping term, the spin vector relaxes to the
direction of $\uvec{e}_\parallel$ as time progresses. The same behavior is
noticed in Figure 6b for the thin film, although the precessional trajectory
is a highly distorted one due to the shape anisotropy.

\begin{figure}[h]
\centering\includegraphics[width=1.07\linewidth]{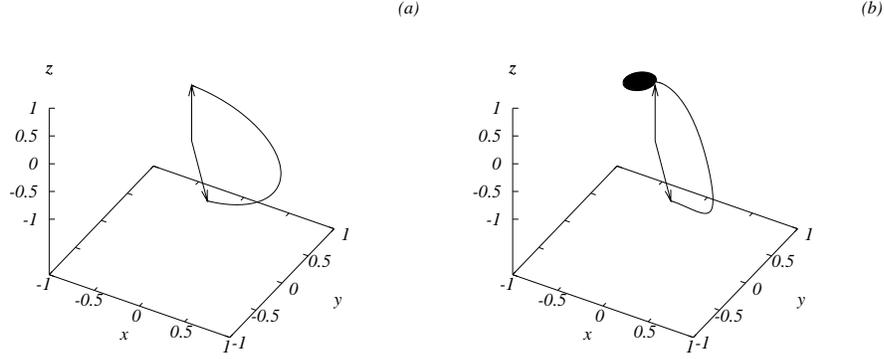}
\caption{Evolution of the magnetization vector ${\bf S}$ in a period of $25 ns$
induced by the spin-polarized current pulse in Figure 4,
(a) with and (b) without anisotropy for the isotropic sample, with 
$N_1=N_2=N_3=1/3$ all other parameters remaining same. A nonzero damping is
assumed in both cases. The current pulse acts
on the magnetization vector for the first $5~ns$. In both cases switching 
happens in the first $5~ns$. In the presence of nonzero 
anisotropy field, (b), the 
magnetization vector precesses to the fixed point near $\uvec{z}$. }
\end{figure}
\begin{figure}[h]
\centering\includegraphics[width=1.07\linewidth]{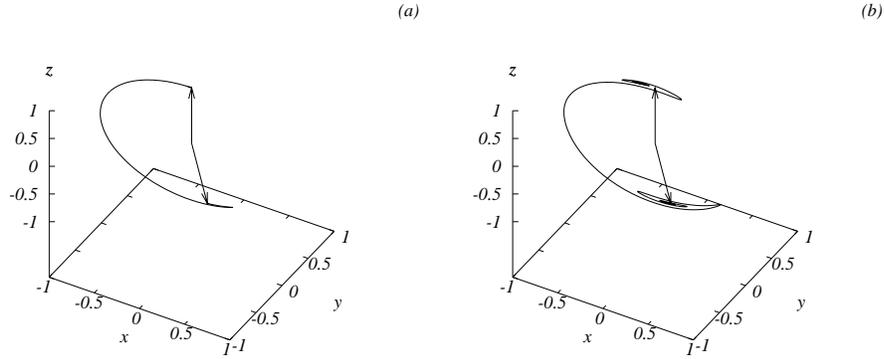}
\caption{Evolution of the magnetization vector ${\bf S}$ in a period of $25 ns$
induced by the spin-polarized current pulse in Figure 4,
for a infinite thin film sample, with $N_1=0=N_3$, and $N_2=1$, all other 
parameters remaining same, along with a nonzero damping. (a) Without 
anisotropy and (b) with anisotropy. As in Figure 5, switching happens in
the first $5~ns$. }
\end{figure}
\section{Discussion and conclusion}
We have shown using analytical study and numerical analysis of the nonlinear
dynamics underlying the magnetization behavior in spin-valve pillars 
that a very effective switching of
macro-magnetization vector can be achieved by a spin 
transfer-torque, modeled using an extended LL equation. 
Rewriting the extended LL equation using
the complex stereographic variable, we find the spin-transfer torque
term indeed acts as an imaginary applied field term, and can lead to both
precession and dissipation. It has also been pointed out why the spin-torque
term is more effective in switching the magnetization vector compared to
the applied field. On application of a spin-polarized current the average 
magnetization vector in the free layer was shown to switch to the direction of 
polarization of the spin polarized current. For a constant current, the
required current density was found to be of the order of $10^8A/cm^2$.
For recording in magnetic media, switching is achieved using a stronger 
polarized current pulse of order $10^9A/cm^2$. Currents of these magnitudes
have been achieved experimentally.

\ack
The work forms part of a research project sponsored by the Department of 
Science
and Technology, Government of India and a DST Ramanna Fellowship to M. L. 


\end{document}